\documentclass[twocolumn,aps,prl,groupedaddress]{revtex4-1}
\usepackage{bm}
\usepackage{amsfonts}
\usepackage{amsmath}
\usepackage{amssymb}
\usepackage{geometry}
\usepackage{graphicx}
\usepackage{footmisc}
\usepackage{braket}
\usepackage{array}
\usepackage{threeparttable}
\usepackage{bibentry,natbib}
\usepackage{bigstrut}
\usepackage[colorlinks,linkcolor=blue,anchorcolor=blue,citecolor=blue,urlcolor=blue]{hyperref}

\begin{document}

\title{Precision calculation of hyperfine structure and the Zemach radii of $^{6,7}$Li$^+$ ions}

\author{Xiao-Qiu Qi $^{1,4,*}$}
\author{Pei-Pei Zhang $^{1,3,*}$~\footnotetext{*These authors contributed equally to this work.}}
\author{Zong-Chao Yan $^{2,1,5}$}
\author{G. W. F. Drake $^{3}$}
\author{Zhen-Xiang Zhong $^{1,\dag}$~\footnotetext{$\dag$Email address: zxzhong@wipm.ac.cn}}
\author{Ting-Yun Shi $^{1,5}$ }
\author{\;\;\; Shao-Long Chen $^{1}$ }
\author{\;\;\; Yao Huang $^{1}$ }
\author{\;\;\; Hua Guan $^{1,\ddag}$~\footnotetext{$\ddag$Email address: guanhua@wipm.ac.cn}}
\author{\;\;\; Ke-Lin Gao $^{1,5}$ }

\affiliation {$^1$ State Key Laboratory of Magnetic Resonance and Atomic and Molecular Physics, Wuhan Institute of Physics and Mathematics, Innovation Academy for Precision Measurement Science and Technology, Chinese Academy of Sciences, Wuhan 430071, China}
\affiliation {$^2$ Department of Physics, University of New Brunswick, Fredericton, New Brunswick, Canada E3B 5A3}
\affiliation {$^3$ Department of Physics, University of Windsor, Windsor, Ontario, Canada N9B 3P4}
\affiliation {$^4$ University of Chinese Academy of Sciences, Beijing 100049, China}
\affiliation {$^5$ Center for Cold Atom Physics, Chinese Academy of Sciences, Wuhan 430071, China}

\begin{abstract}
 The hyperfine structures of the $2\,^3\!S_1$ states of the $^6$Li$^+$ and $^7$Li$^+$ ions are investigated theoretically to extract the Zemach radii of the $^6$Li and $^7$Li nuclei by comparing with precision measurements. The obtained Zemach radii are larger than the previous values of Puchalski and Pachucki [\href{https://link.aps.org/doi/10.1103/PhysRevLett.111.243001}{Phys. Rev. Lett. {\bf 111}, 243001 (2013)}] and disagree with them by about 1.5 and 2.2 standard deviations for $^6$Li and $^7$Li, respectively. Furthermore, our Zemach radius of $^6$Li differs significantly from the nuclear physics value, derived from the nuclear charge and magnetic radii [\href{https://link.aps.org/doi/10.1103/PhysRevA.78.012513}{Phys. Rev. A {\bf 78}, 012513 (2008)}], by more than 6 sigma, indicating an anomalous nuclear structure for $^6$Li. The conclusion that the Zemach radius of $^7$Li is about 40\% larger than that of $^6$Li is confirmed. The obtained Zemach radii are used to calculate the hyperfine splittings of the $2\,^3\!P_J$ states of $^{6,7}$Li$^+$, where an order of magnitude improvement over the previous theory has been achieved for $^7$Li$^+$.
\end{abstract}
\date{\today}
\maketitle

\emph{Introduction.}---
High precision atomic physics measurements \cite{Pastor2012,Luo2013,Zheng2017,Kato2018,Rengelink2018,Schmidt2018,Thomas2020} and associated theory \cite{Yan1995,Pachucki2006L,Pachucki2010} are playing a rapidly increasing role as a probe for both nuclear structure and new physics.  In addition to helium, the Li$^+$ ion is a promising candidate to probe the distribution of magnetic moment inside the nucleus as characterized by the Zemach radius \cite{Zemach1956}.  Recent work on hyperfine structure (HFS) for the ground state of neutral lithium \cite{Puchalski2013} suggested a large unexplained discrepancy for the Zemach radius of $^6$Li, as opposed to relatively good agreement for $^7$Li. Their results showed that, although the nuclear charge radius of $^7$Li is smaller than $^6$Li, the Zemach radius is about 40\% larger than $^6$Li, which is inconsistent with the nuclear data value, as cited by Yerokhin \cite{Yerokhin2008}. The analysis depends critically on the theory of hyperfine structure for the isotopes $^6$Li and $^7$Li, including higher-order relativistic and quantum electrodynamic (QED) effects up to the $m\alpha^6$ limit of current technology.  From this, one can determine the Zemach radius as a variable parameter in comparing theory with experiment.

The present work is motivated by an analysis of the recent experimental results for the HFS of $^7$Li$^+$ of Guan {\it et al.}\ \cite{Guan2020}, as well as by the ongoing experiment on $^6$Li$^+$ in our laboratory at Wuhan.  Following earlier work \cite{Schuler1924,Fan1978,Kotz1981,Kowalski1983,Riis1994,Clarke2003}, their measurements represent a major step forward in precision for the fine and hyperfine structure for the $2\,^3\!S_1$ and $2\,^3\!P_J$ states of $^7$Li$^+$, with uncertainties of less than 100 kHz. These experimental activities will present an important opportunity to probe the nuclear structure of lithium isotopes, particularly for the existing large discrepancy between theory and experiment in the Zemach radius of the $^6$Li nucleus. We calculate here the hyperfine splittings of the $2\,^3\!S_1$ and $2\,^3\!P_J$ states of $^6$Li$^+$ and $^7$Li$^+$ with QED corrections included up to order $m\alpha^6$, and find a similarly large disagreement with experiment for the Zemach radius of $^6$Li.

The advantage of working with the Li$^+$ ion is that it is a two-electron system for which highly accurate nonrelativistic wave functions in Hylleraas coordinates are available \cite{ZhangPeiPei2015}, and so this is removed as a source of uncertainty for all practical purposes. The relativistic corrections of order $m\alpha^4$ to the HFS of $^{6,7}$Li$^+$ were calculated by Drake {\it et al.}~\cite{Riis1994}, including the contributions from the nuclear electric quadrupole moment and other nuclear structure effects, and the theoretical accuracy is sub-MHz and MHz, respectively. For the parallel two-electron case of $^3$He, the hyperfine structures of the $2\,^3\!S_1$ and $2\,^3\!P_J$ states were studied by Pachucki {\it et al.}~\cite{Pachucki2001,Pachucki2012}, including the QED corrections up to order $m\alpha^6$.  Their calculations for $2\,^3\!P_J$ improved the previous theoretical predictions by an order of magnitude.

\begin{figure}[ht]
\includegraphics[scale=0.7]{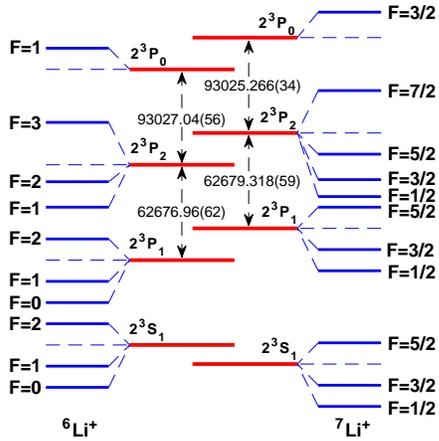}
\caption{Hyperfine energy levels ( not drawn to scale ) of the $2\,^3\!S_1$ and $2\,^3\!P_J$ states of $^{6,7}$Li$^+$~\cite{Riis1994,Pachucki2010}, in MHz.}
\label{fig:1}
\end{figure}

\emph{Theoretical method.}---The nonrelativistic quantum electrodynamic (NRQED) theory for quasidegenerate states is used to calculate fine and hyperfine structure splittings~\cite{Puchalski2009,Pachucki2012,Pachucki2019,Haidar2020}. In order to obtain the energies of the $2\,^3\!\chi^F_J$ ($\chi=S$ or $P$) states, where the energy level diagram is shown in Figure~\ref{fig:1}, we need to diagonalize the effective Hamiltonian $H$
with its matrix elements being
\begin{equation}\label{eq:1}
\begin{aligned}
E_{JJ'}^F\equiv \langle JFM_F|H|J'FM_F \rangle,
\end{aligned}
\end{equation}
where $M_F$ is the projection of the total angular momentum $F$, which can be fixed arbitrarily since the energies are independent of it. For convenience, we treat the $2\,^3P_J$ centroid as a zero level. The above matrix elements (\ref{eq:1}) can be expanded in powers of the fine structure constant $\alpha$
\begin{equation}\label{eq:2}
    \begin{aligned}
    E_{JJ'}^F  =&\langle H_{\mathrm{fs}}\rangle_J \delta_{JJ'} + \langle H_{\mathrm{hfs}}^{(4+)}\rangle + \langle H_{\mathrm{hfs}}^{(6)}\rangle \\
    &+ 2\langle H_{\mathrm{hfs}}^{(4)} , [H_{\mathrm{nfs}}^{(4)}+H_{\mathrm{fs}}^{(4)}] \rangle  + \langle H_{\mathrm{hfs}}^{(4)} , H_{\mathrm{hfs}}^{(4)} \rangle \\
    &+ \langle H_{\mathrm{QED}}^{(6)}\rangle + \langle H_{\mathrm{QED}}^{\mathrm{ho}} \rangle + \langle H_{\mathrm{nucl}} \rangle + \langle H_{\mathrm{eqm}} \rangle,
    \end{aligned}
\end{equation}
where $\langle A,B \rangle \equiv \langle A \frac{1}{(E_0-H_0)'} B \rangle$, with $H_0$ and $E_0$ being the nonrelativistic Hamiltonian and its eigenvalue. $H_{\mathrm{fs}}$ is the effective operator that does not depend on the nuclear spin and is responsible for the fine structure splittings~\cite{Pachucki2010,Riis1994}. The other terms in Eq.~(\ref{eq:2}) are the nuclear spin dependent contributions. $H_{\mathrm{hfs}}^{(4+)}$ is the leading-order hyperfine Hamiltonian of $m\alpha^4$, where the superscript `+' means the higher-order terms from the recoil and anomalous magnetic moment effects. $H_{\mathrm{hfs}}^{(6)}$ is the effective operator for the hyperfine  splittings of order $m\alpha^6$. $H_{\mathrm{fs}}^{(4)}$ and $H_{\mathrm{nfs}}^{(4)}$ are the Breit Hamiltonians of order $m\alpha^4$ with and without electron spin. The fifth term in Eq.~(\ref{eq:2}) is the second-order hyperfine correction, which contributes to the isotope shift, fine and hyperfine splittings. $H_{\mathrm{QED}}^{(6)}$ and $H_{\mathrm{QED}}^{\mathrm{ho}}$ are the two effective operators for the QED corrections of order $m\alpha^6$ and higher $\sim m\alpha^7$. Finally, $H_{\mathrm{nucl}}$ and $H_{\mathrm{eqm}}$ represent the nuclear effects due to the Zemach radius and the nuclear electric quadrupole moment. The detailed forms of these operators are given in Sections I and II of the Supplemental Material.

\begin{table*}[ht]
    \scriptsize\caption{\label{tab:1} The Zemach radii of $^6$Li$^+$ and $^7$Li$^+$. For each ion, two values of the Zemach radius are extracted based on two different transitions, in MHz. In the table, $E_{\mathrm{theor}}$ represents the theoretical value of the HFS without the contribution of nuclear structure and $R_{\rm e}$ is the nuclear charge radius.}
     \begin{ruledtabular}
     \begin{tabular}{lllllllll}
     \multicolumn{1}{c}{$ $}
    &\multicolumn{2}{c}{$^6$Li$^+$}
    &\multicolumn{2}{c}{$^7$Li$^+$} \\
    \cline{2-3} \cline{4-5}
    &$2\,^3\!S_1^0-2\,^3\!S_1^1$   &$2\,^3\!S_1^1-2\,^3\!S_1^2$   &$2\,^3\!S_1^{1/2}-2\,^3\!S_1^{3/2}$   &$2\,^3\!S_1^{3/2}-2\,^3\!S_1^{5/2}$             \\
    \hline
    $E_{\mathrm{theor}}$                                        &3002.597(22)         &6005.279(14)           &11894.581(69)        &19825.291(46)        \\
    $E_{\mathrm{expt}}$, Kowalski \emph{et al.}~\cite{Kowalski1983} &3001.780(50)     &6003.600(50)           &                     &                     \\
    $E_{\mathrm{expt}}$, Guan \emph{et al.}~\cite{Guan2020}     &                     &                       &11890.088(65)        &19817.696(42)        \\
    $(E_{\mathrm{expt}}-E_{\mathrm{theor}})/E_{\mathrm{expt}}$  &--272(18) ppm        &--280(9) ppm           &--378(8) ppm          &--383(3) ppm        \\
    Puchalski \emph{et al.} ~\cite{Puchalski2013}               &\multicolumn{2}{c}{--261(3) ppm}             &\multicolumn{2}{c}{--368(3) ppm}            \\
    Yerokhin ~\cite{Yerokhin2008}                               &\multicolumn{2}{c}{\,\,\,--368(60) ppm}      &\multicolumn{2}{c}{\,\,\,--369(23) ppm}     \\
    Li \emph{et al.} ~\cite{Haibin2020,Puchalski2013}           &\multicolumn{2}{c}{--277(7) ppm}             &\multicolumn{2}{c}{ }                       \\
    $R_{\mathrm{em}}$, this work                                &2.40(16) fm          &2.47(8) fm             &3.33(7) fm           &3.38(3) fm            \\
    $R_{\mathrm{em}}$, Puchalski \emph{et al.}~\cite{Puchalski2013} &\multicolumn{2}{c}{2.30(3) fm}           &\multicolumn{2}{c}{3.25(3) fm}              \\
    $R_{\mathrm{em}}$, Nuclear physics~\cite{Yerokhin2008}          &\multicolumn{2}{c}{\,\,\,3.71(16) fm}     &\multicolumn{2}{c}{3.42(6) fm}       \\
    $R_{\mathrm{em}}$, Li \emph{et al.} ~\cite{Haibin2020,Puchalski2013}  &\multicolumn{2}{c}{2.44(6) fm}     &\multicolumn{2}{c}{ }                       \\
    $R_{\rm e}$, Lu \emph{et al.}~\cite{Drake2013}              &\multicolumn{2}{c}{\quad2.589(39) fm}        &\multicolumn{2}{c}{\quad2.444(44) fm}       \\
    \end{tabular}
    \end{ruledtabular}
\end{table*}

In our calculations, we use the Hylleraas variational technique~\cite{ZhangPeiPei2015} to determine high-precision wave functions corresponding to the nonrelativistic part of the Hamiltonian, and then calculate the relativistic, QED, and nuclear effects order by order.
Two different wave functions with and without the mass polarization term in the nonrelativistic Hamiltonian are generated.
In calculating the second-order terms of $m\alpha^6$, the coupling of intermediate states with different symmetries should be included, where some singular terms are treated by including more singular functions in the intermediate states~\cite{Drake2002}. All the operators in this work can be expressed in terms of the following ten basic angular momentum operators~\cite{Pachucki2012}: $S^iL^i$, $I^iL^i$, $I^iS^i$, $\{S^iS^j\}\{L^iL^j\}$, $I^iS^j\{L^iL^j\}$, $I^iL^j\{S^iS^j\}$, $\{I^iI^j\}\{S^iS^j\}$, $\{I^iI^j\}\{L^iL^j\}$, $S^iL^j\{I^iI^j\}$, and $\{I^iI^j\}\{\{S^mS^n\}\{L^kL^l\}\}^{ij}$, where $ S^iL^i\equiv  \vec{S} \cdot \vec{L} $, $\{S^iS^j\}$ is the second-order tensor part defined by $\{S^iS^j\}\equiv \frac{1}{2}S^iS^j + \frac{1}{2}S^jS^i - \frac{1}{3}\vec{S}^2\delta^{ij}$, and the summation over the repeated indices is assumed. The matrix elements of these operators can be evaluated analytically using Racah algebra.

\begin{figure}[ht]
\includegraphics[width=0.5\textwidth]{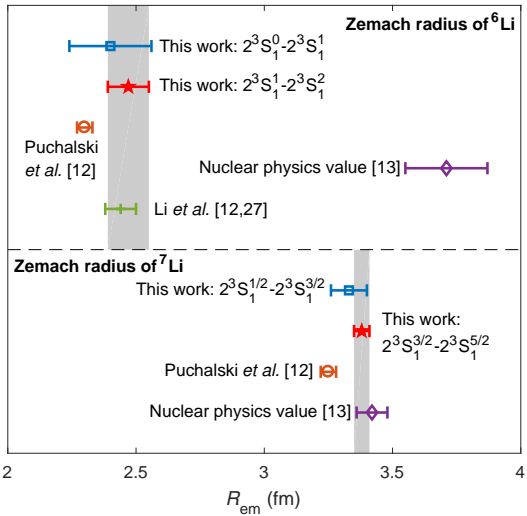}
\caption{Comparison of the Zemach radii of $^6$Li and $^7$Li.}
\label{fig:2}
\end{figure}

\emph{Zemach radii.}---A combination of the experimental~\cite{Kowalski1983,Guan2020} and theoretical results for the HFS of the $2\,^3\!S_1$ state can be used to determine the contribution of nuclear structure and extract the Zemach radii. Numerical values of the relevant operators are presented in Section III of the Supplemental Material. Our results are given in Table~\ref{tab:1}. Since the theoretical uncertainties of our calculations are mainly from the order $m\alpha^7$ term $H_{\rm QED}^{\rm ho}$, they are taken to be 10\% of this contribution calculated approximately, see the Supplementary Material. Two determinations of the Zemach radii from the $^{6,7}$Li$^+$ ions are obtained independently based on two different transitions, which are in good agreement with each other. The uncertainty of the Zemach radius from the $2\,^3\!S_1^0-2\,^3\!S_1^1$ transition of $^6$Li$^+$ is larger, which is caused by the accuracy of experimental measurements. For $^7$Li$^+$, we also combined our theoretical calculations with the experimental values from Kowalski \emph{et al.}~\cite{Kowalski1983} and derived the two Zemach radii $3.38(6)$~fm and $3.39(3)$~fm, which are consistent with those in Table~\ref{tab:1} extracted from the measurements of Guan {\it et al.}~\cite{Guan2020}. Thus, we choose $2.47(8)$~fm for $^6$Li and $3.38(3)$~fm for $^7$Li as the final recommended values of the Zemach radii. An important feature to note is that, to a good approximation, $(E_{\mathrm{expt}}-E_{\mathrm{theor}})/E_{\mathrm{expt}}$  for an $S$ state is directly related to the nuclear Zemach
radius, where
$E_{\mathrm{theor}}$ is the theoretical value without inclusion of the nuclear term, {\it i.e.},
\begin{equation}\label{eq:3}
\begin{aligned}
\frac{E_{\mathrm{expt}}-E_{\mathrm{theor}}}{E_{\mathrm{expt}}}=\frac{-2ZR_{\mathrm{em}}}{a_0}\,.
\end{aligned}
\end{equation}
In other words, this relation is valid for both neutral $^{6,7}$Li and ionic $^{6,7}$Li$^+$. Also in the table are the values derived by Yerokhin from the nuclear charge and magnetic radii~\cite{Yerokhin2008}. Figure~\ref{fig:2} shows a comparison of the resulting Zemach radii. Our Zemach radius for $^7$Li agrees with the nuclear physics value, whereas our result for $^6$Li disagrees by $6\sigma$ (standard deviations) from the value of 3.71(16) fm. Furthermore, our values are all larger than those of Puchalski and Pachucki~\cite{Puchalski2013} by $1.5\sigma$ and $2.2\sigma$ for $^6$Li and $^7$Li, respectively. However, their Zemach radii were extracted using the experimental values of Beckmann \emph{et al.}~\cite{Beckmann1974}. If instead we combine the most recent measurement of $^6$Li by Li \emph{et al.}~\cite{Haibin2020} with the calculation of Puchalski and Pachucki, the Zemach radius of $^6$Li turns out to be 2.44(6) fm, which is in agreement with our result, as shown in Table~\ref{tab:1} and Figure~\ref{fig:2}. Our results confirm the conclusion that the Zemach radius of $^7$Li is about 40\% larger than that of $^6$Li, as was first pointed out by Puchalski and Pachucki~\cite{Puchalski2013}.

\begin{table}[ht]
    \scriptsize\caption{\label{tab:2} Theoretical results for individual $2\,^3\!P^F_J$ levels in $^{6,7}$Li$^+$, relative to the $2\,^3\!P_J$ centroid energy, where the first error in each entry is due to the fine structure and the second error is due to the hyperfine structure, in MHz. The Zemach radius used is $2.47(8)$~fm for $^6\mathrm{Li}^+$ and $3.38(3)$~fm for $^7\mathrm{Li}^+$.}
     \begin{ruledtabular}
         \begin{tabular}{lccc}
     \multicolumn{1}{c}{$ $}
    &\multicolumn{1}{c}{$^6\mathrm{Li}^+$}
    &\multicolumn{1}{c}{$ $}
    &\multicolumn{1}{c}{$^7\mathrm{Li}^+$} \\
    \hline
    $2^3P_0^{F=1}$     &103646.261(612)(1)           &$2^3P_0^{F=3/2}$   &104395.489(54)(6)    \\
    $2^3P_2^{F=3}$     &13371.488(283)(27)           &$2^3P_2^{F=7/2}$   &21709.266(27)(53)    \\
    $2^3P_2^{F=2}$     &9243.564(283)(14)            &$2^3P_2^{F=5/2}$   &9936.263(27)(15)    \\
    $2^3P_2^{F=1}$     &6385.602(283)(40)            &$2^3P_2^{F=3/2}$   &327.952(27)(52)      \\
    $2^3P_1^{F=2}$     &--50782.508(350)(14)         &$2^3P_2^{F=1/2}$   &--5875.456(27)(79)    \\
    $2^3P_1^{F=1}$     &--53670.888(350)(14)         &$2^3P_1^{F=5/2}$   &--47679.845(33)(27)    \\
    $2^3P_1^{F=0}$     &--54988.619(350)(27)         &$2^3P_1^{F=3/2}$   &--57646.278(33)(20)    \\
    $            $     &                             &$2^3P_1^{F=1/2}$   &--61885.185(33)(45)    \\
    \end{tabular}
    \end{ruledtabular}
\end{table}

\begin{table}[ht]
    \scriptsize\caption{\label{tab:3} Theoretical hyperfine transitions in the $2\,^3\!P_J$ states of $^6$Li$^+$, in MHz. In our work, the nuclear
electric quadrupole moment used is $Q_d=-0.000806(6) \times 10^{-24} \; \mathrm{cm}^2$~\cite{Stone2016} and the Zemach radius is $R_{\mathrm{em}}=2.47(8)$ fm.}
     \begin{ruledtabular}
     \begin{tabular}{lccc}
    $\rm{State}$  &$(J,F)-(J',F')$ &Drake {\it et al.}~\cite{Riis1994}   & This work    \\
    \hline
    $2\,^3\!P_2$  &$(2,1)-(2,2)$   &2858.002(61)    &2857.962(43)  \\
    $$            &$(2,2)-(2,3)$   &4127.882(44)    &4127.924(31)  \\
    $2\,^3\!P_1$  &$(1,0)-(1,1)$   &1317.649(47)    &1317.732(31)  \\
    $$            &$(1,1)-(1,2)$   &2888.327(29)    &2888.379(20)  \\
    \end{tabular}
    \end{ruledtabular}
\end{table}

\begin{table*}[ht]
    \scriptsize\caption{\label{tab:4} Experimental and theoretical hyperfine transitions in the $2\,^3\!P_J$ states of $^7$Li$^+$, in MHz. In our work, the nuclear electric quadrupole moment used is $Q_d=-0.0400(3)\times10^{-24}$ $\rm{cm}^2$~\cite{Stone2016} and the Zemach radius is $R_{\mathrm{em}}=3.38(3)$ fm.}
     \begin{ruledtabular}
     \begin{tabular}{ccccccccc}
     \multicolumn{1}{c}{}
    &\multicolumn{1}{c}{}
    &\multicolumn{3}{c}{Experiment}
    &\multicolumn{2}{c}{Theory}\\
    \cline{3-5} \cline{6-7}
     $\rm{State}$       &$(J,F)-(J',F')$    &K{\"o}tz {\it et al.}~\cite{Kotz1981,Kowalski1983} &Clarke {\it et al.}~\cite{Clarke2003}  & Guan {\it et al.} ~\cite{Guan2020} &Drake {\it et al.}~\cite{Riis1994}   & This work  \\
    \hline
     $2\,^3\!P_2$  &$(2,1/2)-(2,3/2)$  &6203.6(5)  &6204.52(80) &6203.319(67) &6203.27(30)  &6203.408(95)  \\
     $$            &$(2,3/2)-(2,5/2)$  &9608.7(20) &9608.90(49) &9608.220(54) &9608.12(15)  &9608.311(54)  \\
     $$            &$(2,5/2)-(2,7/2)$  &11775.8(5) &11774.04(94)&11772.965(74)&11773.05(18) &11773.003(55) \\
     $2\,^3\!P_1$  &$(1,1/2)-(1,3/2)$  &4237.8(10) &4239.11(54) &4238.823(111)&4238.86(20)  &4238.920(49)  \\
     $$            &$(1,3/2)-(1,5/2)$  &9965.2(6)  &9966.30(69) &9966.655(102)&9966.14(13)  &9966.444(34)  \\
    \end{tabular}
    \end{ruledtabular}
\end{table*}

\begin{figure*}[ht]
\includegraphics[scale=0.7]{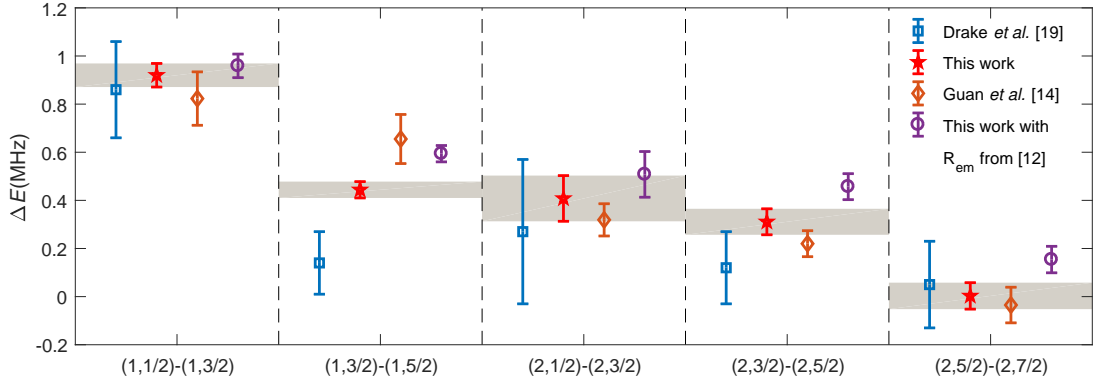}
\caption{Comparison of HFS in the $2\,^3\!P_J$ states of $^7$Li$^+$. $\Delta E$ stands for our results for the five indicated transitions relative to 4238, 9966, 6203, 9608, and 11773, respectively. The blue, red, and purple lines represent the calculations of Drake \emph{et al.}~\cite{Riis1994}, the present calculations using our $R_{\mathrm{em}}=3.38(3)$~fm, and the present calculations using the value of Puchalski and Pachucki~\cite{Puchalski2013} $R_{\mathrm{em}}=3.25(3)$~fm, respectively. The brown lines are the experimental results of Guan \emph{et al.}~\cite{Guan2020}.}
\label{fig:3}
\end{figure*}

\emph{HFS of $2\,^3\!P_J$.}---We calculate the HFS of the $2\,^3\!P_J$ states using our obtained Zemach radii. Numerical values of the relevant operators are shown in Section IV of the Supplemental Material. Since the contribution from the $1s$ electron dominates higher-order QED correction, the assumption $H_{\mathrm{QED}}^{\mathrm{ho}} (1s2p) \simeq H_{\mathrm{QED}}^{\mathrm{ho}} (1s)$ is used. The uncertainty of this correction is also estimated as 10\% of its contribution. According to Eq.~(\ref{eq:2}), the HFS calculations of the $2\,^3\!P_J$ states require the results of the fine structure splittings, which are $\langle H_{fs}\rangle_{J=0}=(8f_{01}+5f_{12})/9$, $\langle H_{fs}\rangle_{J=1}=(-f_{01}+5f_{12})/9$, and $\langle H_{fs}\rangle_{J=2}=(-f_{01}-4f_{12})/9$, relative to the $2\,^3\!P_J$ centroid, with $f_{01}=155704.00(57)$~MHz and $f_{12}=-62676.96(62)$~MHz for $^6$Li$^+$~\cite{Riis1994} and $f_{01} =155704.584(48)$~MHz and $f_{12} =-62679.318(59)$~MHz for $^7$Li$^+$~\cite{Pachucki2010}. The HFS of $2\,^3\!P_J$ of $^6$Li$^+$ and $^7$Li$^+$ can be obtained by diagonalizing the matrix in Eq.~(\ref{eq:2}) and the results relative to the $2\,^3\!P_J$ centroid are listed in Table~\ref{tab:2}.

In calculating the second-order energy, we subtract out the dominant $2\,^1\!P_1 - 2\,^3\!P_1$ singlet-triplet mixing term (ignoring for now hyperfine structure for purposes of illustration) and replace it with the energy shift $\Delta$ obtained by exact diagonalization of the corresponding $2\times2$ Hamiltonian matrix, thereby summing the perturbation series for this term to infinity \cite{Drake1979,Morton2017,Wienczek2019} according to the formula
\begin{equation}\label{eq:4}
    \begin{aligned}
    \tilde{E} = E - \frac{|\langle 2\,^3\!P |H_{\mathrm{mix}}| 2\,^1\!P \rangle|^2}{E(2\,^3\!P)-E(2\,^1\!P)}  + \Delta,
    \end{aligned}
\end{equation}
where $H_{\mathrm{mix}}$  is the singlet-triplet mixing operator. This procedure rapidly becomes essential with increasing $Z$ or with increasing $L$ in order to avoid saturation of singlet-triplet mixing. This modification of the mixing effect alters the hyperfine transitions $(1,1/2)-(1,3/2)$ and $(1,3/2)-(1,5/2)$ of $^7$Li$^+$ by 14(2)~kHz and 11(3)~kHz, respectively. More details are presented in Section V of the Supplemental Material. Our results of HFS for $^6$Li$^+$ and $^7$Li$^+$ are shown in Tables~\ref{tab:3} and \ref{tab:4}. It is noted that the present theoretical values listed in the last column of Table \ref{tab:4} improve the previous corresponding ones in Ref.~\cite{Guan2020} by including the small contributions from the second-order $m\alpha^6$ corrections, as well as by treating the singlet-triplet mixing more rigorously, as mentioned above.

Tables~\ref{tab:3} and \ref{tab:4} show that our results have uncertainties less than 100 kHz. The theoretical uncertainty mainly comes from the Zemach radius and the contribution of $m\alpha^7$. For $^6$Li$^+$, our results are consistent with those of Drake \emph{et al.}~\cite{Riis1994} at the same level of accuracy. For $^7$Li$^+$, the calculations of Drake \emph{et al.}~\cite{Riis1994}
have been improved by about one order of magnitude, with the only exception that the value of Drake \emph{et al.}~\cite{Riis1994} for the $(1,3/2)-(1,5/2)$ interval in $2\,^3\!P_1$ differs from the present calculation. The discrepancy is due to the use of a different value of the nuclear electric quadrupole moment $Q_{d}$, because this interval is particularly sensitive to $Q_{d}$. It is noted that there is a discrepancy between the experimental value of Guan \emph{et al.}~\cite{Guan2020} and our calculation for the interval $(1,3/2)-(1,5/2)$ of $2\,^3\!P_1$ of $^7$Li$^+$, which are only consistent within $1.6\sigma$. We do not have a satisfactory explanation for this discrepancy and we now call for more investigation on the Li$^+$ isotopes. Figure~\ref{fig:3} shows a comparison of HFS transitions of $^7$Li$^+$. From the figure one can see the influence of the Zemach radius on each transition. For example, for the transition $(1,3/2)-(1,5/2)$, the result calculated using the Zemach radius of Puchalski and Pachucki~\cite{Puchalski2013} is in agreement with the measured value~\cite{Guan2020}, but it disagrees with the measured value for the transition $(2,3/2)-(2,5/2)$.

\emph{Conclusion.}---We have studied the HFS of the $2\,^3\!S_1$ and $2\,^3\!P_J$ states of $^6$Li$^+$ and $^7$Li$^+$, including the relativistic and QED corrections up to order $m\alpha^6$. By comparing with the measured HFS of $2\,^3\!S_1$, we have derived the Zemach radii for the $^6$Li and $^7$Li nuclei.  While the result for $^7$Li is in good agreement, the result for $^6$Li disagrees by more than $6\sigma$ from the value derived from the nuclear charge and magnetic radii by Yerokhin \cite{Yerokhin2008}, indicating an anomalous nuclear structure for $^6$Li.  Our results disagree with the previously extracted values from neutral $^{6,7}$Li~\cite{Puchalski2013} by about $1.5\sigma$ and $2.2\sigma$ respectively, but they come into agreement for $^{6}$Li when the more recent measurement of hyperfine structure of $^6$Li by Li \emph{et al.}~\cite{Haibin2020} is used. Our results also confirm the conclusion~\cite{Puchalski2013} that the Zemach radius of $^7$Li is about 40\% larger than that of $^6$Li, even though the charge radius is smaller. Using thus determined Zemach radii, we have calculated the HFS of the $2\,^3\!P_J$ states, where the  $2\,^1\!P_1-2\,^3\!P_J$ mixing has been treated rigorously. Our results for the HFS of $2\,^3\!P_J$ in $^7$Li$^+$ have improved previous calculations by one order of magnitude.

\begin{acknowledgements}
The authors thank X. J. Liu, S. M. Hu, Y. R. Sun and H. P. Liu for helpful discussions. This research was supported by the Strategic Priority Research Program of CAS under Grant No.\ XDB21010400, and by the National Natural Science Foundation of China under Grant Nos.\ 11604369, 91636216, 11974382, 11934014 and 11622434. ZXZ acknowledges the support from the YIPA program of CAS. H. Guan acknowledges the support from CAS Youth Innovation Promotion Association under Grant No.\ Y201963, the Hubei Province Science Fund for Distinguished Young Scholars under Grant No.\ 2017CFA040 and K. C. Wong Education Foundation. Z.-C. Yan and G. W. F. Drake acknowledge the supports from NSERC and SHARCnet of Canada.
\end{acknowledgements}

$*$These authors contributed equally to this work.

$\dag$Email address: zxzhong@wipm.ac.cn

$\ddag$Email address: guanhua@wipm.ac.cn

%

\end{document}


\title{Supplementary Material for\\
``Precision calculation of hyperfine structure and the Zemach radii of $^{6,7}$Li$^+$ ions"}

\author{Xiao-Qiu Qi $^{1,4,*}$}
\author{Pei-Pei Zhang $^{1,3,*}$~\footnotetext{*These authors contributed equally to this work.}}
\author{Zong-Chao Yan $^{2,1,5}$}
\author{G. W. F. Drake $^{3}$}
\author{Zhen-Xiang Zhong $^{1,\dag}$~\footnotetext{$\dag$Email address: zxzhong@wipm.ac.cn}}
\author{Ting-Yun Shi $^{1,5}$ }
\author{\;\;\; Shao-Long Chen $^{1}$ }
\author{\;\;\; Yao Huang $^{1}$ }
\author{\;\;\; Hua Guan $^{1,\ddag}$~\footnotetext{$\ddag$Email address: guanhua@wipm.ac.cn}}
\author{\;\;\; Ke-Lin Gao $^{1,5}$ }

\affiliation {$^1$ State Key Laboratory of Magnetic Resonance and Atomic and Molecular Physics, Wuhan Institute of Physics and Mathematics, Innovation Academy for Precision Measurement Science and Technology, Chinese Academy of Sciences, Wuhan 430071, China}
\affiliation {$^2$ Department of Physics, University of New Brunswick, Fredericton, New Brunswick, Canada E3B 5A3}
\affiliation {$^3$ Department of Physics, University of Windsor, Windsor, Ontario, Canada N9B 3P4}
\affiliation {$^4$ University of Chinese Academy of Sciences, Beijing 100049, China}
\affiliation {$^5$ Center for Cold Atom Physics, Chinese Academy of Sciences, Wuhan 430071, China}

\date{\today}
\maketitle

\section{Relativistic corrections}\label{sec1}

For an atomic system, the relativistic corrections of order $m\alpha^4$ for non-fine-structure, {\it i.e.}, the spin independent part, fine- and hyperfine-structure, as well as the relativistic corrections of order $m\alpha^6$ for hyperfine structure, can be obtained from the Breit Hamiltonian in an external field. The corresponding operators of order $m\alpha^4$ are, respectively, denoted by
$H_{\mathrm{nfs}}^{(4)}$, $H_{\mathrm{fs}}^{(4)}$, and $H_{\mathrm{hfs}}^{(4+)}$, which are given by ~\cite{Pachucki2006A,Pachucki2012,Pachucki2015,Pachucki2019}
\begin{equation}\label{eq:1}
    \begin{aligned}
        H_{\mathrm{nfs}}^{(4)} = G,
    \end{aligned}
\end{equation}
\begin{equation}\label{eq:2}
    \begin{aligned}
        H_{\mathrm{fs}}^{(4)}  &= \vec{S} \cdot \vec{G} + \vec{S}_A \cdot \vec{G}_A + S^iS^j G^{ij}, \\
    \end{aligned}
\end{equation}
\begin{equation}\label{eq:3}
    \begin{aligned}
        H_{\mathrm{hfs}}^{(4+)}=& C^1_{34} \{ (1+a_e) \vec{I} \cdot [ \vec{S} P + \vec{S}_A P_A ] + (1+a_e) I^i [ S^j P^{ij} + S^j_A P_A^{ij} ] + \vec{I} \cdot [\vec{P} + \vec{P}_{\mathrm{rec}}] \}, \\
    \end{aligned}
\end{equation}
where $\vec{S}=(\vec{\sigma}_1 + \vec{\sigma}_2)/2$ and $\vec{S}_A=(\vec{\sigma}_1 - \vec{\sigma}_2)/2$ are the electron spin operators, $a_e$ is the anomalous magnetic moment of the electron, $C^y_{xz}$ denotes $\mu^x [(1+\kappa)/mM]^y \alpha^z$, $m$ and $M$ are the electron and nuclear mass respectively, $\mu=mM/(m+M)$ is the reduced mass, and $\kappa$ is the anomalous magnetic moment of the nucleus. Also, in the above,
\begin{equation}\label{eq:4}
    \begin{aligned}
         G = & -\frac{p^4_1+p^4_2}{8}+\frac{Z \pi}{2}[\delta^3(\vec{r}_1)+ \delta^3(\vec{r}_2)]-\frac{1}{2} p^i_1 \bigg( \frac{\delta^{ij}}{r} + \frac{r^ir^j}{r^3} \bigg) p^j_2, \\
    \end{aligned}
\end{equation}
is the spin-independent part of order $m\alpha^4$, and
\begin{equation}\label{eq:5}
    \begin{aligned}
        \vec{G}=\frac{Z}{4} \bigg( \frac{\vec{r}_1 \times \vec{p}_1}{r_1^3}+\frac{\vec{r}_2 \times \vec{p}_2}{r_2^3} \bigg) + \frac{3}{4} \frac{\vec{r}}{r^3} \times ( \vec{p}_2 - \vec{p}_1 ),
    \end{aligned}
\end{equation}
\begin{equation}\label{eq:6}
    \begin{aligned}
        \vec{G}_A=\frac{Z}{4} \bigg( \frac{\vec{r}_1 \times \vec{p}_1}{r_1^3}-\frac{\vec{r}_2 \times \vec{p}_2}{r_2^3} \bigg) + \frac{1}{4} \frac{\vec{r}}{r^3} \times ( \vec{p}_2 + \vec{p}_1 ),
    \end{aligned}
\end{equation}
and
\begin{equation}\label{eq:7}
    \begin{aligned}
        G^{ij}=\frac{1}{2r^3} \bigg( \delta^{ij} - 3\frac{r^ir^j}{r^2} \bigg),
    \end{aligned}
\end{equation}
are responsible for the fine structure of order $m\alpha^4$, with $Z$ being the nuclear charge.
The operators in Eq.~(\ref{eq:3}) are related to HFS, given by
\begin{equation}\label{eq:8}
\begin{aligned}
P = \frac{4\pi Z}{3} [\delta^3(\vec{r}_1) + \delta^3(\vec{r}_2)],
\end{aligned}
\end{equation}
\begin{equation}\label{eq:9}
\begin{aligned}
P_A = \frac{4\pi Z}{3}  [\delta^3(\vec{r}_1) - \delta^3(\vec{r}_2)],
\end{aligned}
\end{equation}
\begin{equation}\label{eq:10}
\begin{aligned}
\vec{P} = Z \bigg[ \frac{\vec{r}_1 \times \vec{p}_1}{r_1^3} + \frac{\vec{r}_2 \times \vec{p}_2}{r_2^3} \bigg],
\end{aligned}
\end{equation}
\begin{equation}\label{eq:11}
\begin{aligned}
\vec{P}_{\mathrm{rec}}= \frac{m}{M} \frac{1+2\kappa}{1+\kappa} \frac{Z}{2} \bigg[\frac{\vec{r}_1}{r_1^3}+\frac{\vec{r}_2}{r_2^3}\bigg] \times (\vec{p}_1+\vec{p}_2),
\end{aligned}
\end{equation}
\begin{equation}\label{eq:12}
\begin{aligned}
P^{ij} = -\frac{Z}{2}\bigg[\frac{1}{r_1^3} \bigg(\delta^{ij} - 3\frac{r_1^ir_1^j}{r_1^2} \bigg) + \frac{1}{r_2^3} \bigg( \delta^{ij} - 3\frac{r_2^ir_2^j}{r_2^2} \bigg)\bigg],
\end{aligned}
\end{equation}
\begin{equation}\label{eq:13}
\begin{aligned}
P_A^{ij} = -\frac{Z}{2}\bigg[\frac{1}{r_1^3} \bigg(\delta^{ij} - 3\frac{r_1^ir_1^j}{r_1^2} \bigg) - \frac{1}{r_2^3} \bigg( \delta^{ij} - 3\frac{r_2^ir_2^j}{r_2^2} \bigg)\bigg].
\end{aligned}
\end{equation}

The matrix elements of $H_{\mathrm{hfs}}^{(4+)}$ in $2^3\chi^F_J$ ($\chi=S$ or $P$) states are~\cite{Pachucki2012}
\begin{equation}\label{eq:14}
\begin{aligned}
\langle H_{\mathrm{hfs}}^{(4+)}\rangle =& C^1_{34} \bigg[ (1+a_e) \langle P \rangle \langle I^iS^i \rangle + \frac{1}{2} \langle \vec{P} + \vec{P}_{\mathrm{rec}} \rangle \langle I^iL^i \rangle - \frac{3}{5} (1+a_e) \langle \hat{P} \rangle \langle I^iS^j\{L^iL^j\} \rangle \bigg]\,,
\end{aligned}
\end{equation}
where the matrix elements of the three operators
\begin{equation}\label{eq:15}
\begin{aligned}
\langle P \rangle = \langle 2\,^3\!\vec{\chi} |P| 2\,^3\!\vec{\chi} \rangle,
\end{aligned}
\end{equation}
\begin{equation}\label{eq:16}
\begin{aligned}
\langle \vec{P} \rangle = \langle 2\,^3\!\vec{\chi} |\vec{P}| 2\,^3\!\vec{\chi} \rangle,
\end{aligned}
\end{equation}
\begin{equation}\label{eq:17}
\begin{aligned}
\langle \hat{P} \rangle = \langle 2\,^3\!\vec{\chi} |P^{ij}| 2\,^3\!\vec{\chi} \rangle,
\end{aligned}
\end{equation}
are defined in Ref.~\cite{Pachucki2012}. The form of $H_{\mathrm{hfs}}^{(6)}$ is
\begin{equation}\label{eq:18}
    \begin{aligned}
        H_{\mathrm{hfs}}^{(6)} &= C^1_{36} \cdot \big(\vec{I} \cdot \vec{S} K + \vec{I} \cdot \vec{K} + I^iS^j K^{ij} \big), \\
    \end{aligned}
\end{equation}
where
\begin{equation}\label{eq:19}
    \begin{aligned}
        K  = \frac{Z^2}{3r^4_1}  - \frac{4\pi Z}{3} p_1^2  \delta^3(\vec{r}_1) - Z \frac{\vec{r}_1}{r_1^3} \cdot \frac{\vec{r}}{r^3},
    \end{aligned}
\end{equation}
\begin{equation}\label{eq:20}
    \begin{aligned}
        \vec{K}  = - Z \bigg[ p_1^2 \frac{\vec{r}_1 \times \vec{p}_1}{r_1^3} + \frac{\vec{r}_1 \times \vec{p}_2}{rr_1^3} + \bigg( \frac{\vec{r}_1}{r_1^3} \times \frac{\vec{r}}{r^3} \bigg)(\vec{r} \cdot \vec{p}_2) \bigg],
    \end{aligned}
\end{equation}
\begin{equation}\label{eq:21}
    \begin{aligned}
        K^{ij}  =& \frac{Z}{2} \bigg[\bigg( \frac{Z}{3r_1} + p_1^2 \bigg)\frac{1}{r_1^3} \bigg( \delta^{ij} - 3\frac{r_1^ir_1^j}{r_1^2} \bigg) + 3\frac{r^i_1}{r_1^3} \frac{r^j}{r^3}-\delta^{ij}\frac{\vec{r}_1}{r_1^3} \cdot \frac{\vec{r}}{r^3} \bigg]. \\
    \end{aligned}
\end{equation}
Though the operator $K$ is singular, its divergent part can be cancelled with the second-order $m\alpha^6$ contribution
\begin{equation}\label{eq:22}
    \begin{aligned}
        E^{(6)}_{\mathrm{sing}}=C^1_{36} \vec{I} \cdot \vec{S} [ \langle K \rangle + \langle P,G \rangle + \langle G,P \rangle ]\,,
    \end{aligned}
\end{equation}
where $\langle P,G\rangle\equiv\langle P\frac{1}{(E_0-H_0)'} G\rangle$, with $H_0$ and $E_0$ being the nonrelativistic Hamiltonian and its eigenvalue.
The operators $P$ and $G$ are both finite, but when combined into a second-order perturbation, they turn out to be divergent. Using the transformations $P' = P+\frac{2}{3} T$ and $G' = G-\frac{1}{4} T$, we have
\begin{equation}\label{eq:23}
\begin{aligned}
\langle K \rangle + \langle P,G \rangle + \langle G,P \rangle=\langle K' \rangle + \langle P',G' \rangle + \langle G',P' \rangle,
\end{aligned}
\end{equation}
where $T \equiv \sum_a\{\frac{Z}{r_a},E_0-H_0\}$. Through these transformations, $\langle K' \rangle$, $\langle P',G' \rangle$, and $\langle G',P' \rangle$ are all finite:
\begin{equation}\label{eq:24}
\begin{aligned}
\langle P' \rangle=-\frac{2Z}{3}\bigg\langle \frac{\vec{r}_1}{r^3_1}\cdot\nabla_1 +\frac{\vec{r}_2}{r^3_2}\cdot\nabla_2 \bigg\rangle,
\end{aligned}
\end{equation}
\begin{widetext}
\begin{equation}\label{eq:25}
\begin{aligned}
\langle G' \rangle =&-\frac{1}{2} \bigg\langle \bigg(E+\frac{Z}{r_1}+\frac{Z}{r_2}-\frac{1}{r}\bigg)^2-\frac{1}{2}\nabla^2_1\nabla^2_2+p_1^i\bigg(\frac{\delta^{ij}}{r}+\frac{r^ir^j}{r^3}\bigg)p^j_2
+\frac{Z}{2}\frac{\vec{r}_1}{r^3_1}\cdot\nabla_1+\frac{Z}{2}\frac{\vec{r}_2}{r^3_2}\cdot\nabla_2-\frac{\vec{r}}{r^3}\cdot(\nabla_1-\nabla_2) \bigg\rangle,
\end{aligned}
\end{equation}
\begin{equation}\label{eq:26}
\begin{aligned}
\langle K' \rangle &=\frac{2}{3} \bigg \langle \bigg(E-\frac{1}{r}\bigg)^2 \bigg(\frac{Z}{r_1}+\frac{Z}{r_2}\bigg) + \bigg( E-\frac{1}{r} \bigg)\bigg(\frac{Z^2}{r^2_1} + \frac{Z^2}{r^2_2} + 4\frac{Z}{r_1}\frac{Z}{r_2} \bigg)+ 2\frac{Z}{r_1}\frac{Z}{r_2}\bigg(\frac{Z}{r_1}+ \frac{Z}{r_2} \bigg)+ p^i_1 \frac{Z^2}{r^2_1} p^i_1  +  p^2_2 \frac{Z^2}{r^2_1} \\
&- p^2_2 \frac{Z}{r_1} p^2_1-\bigg(E-\frac{1}{r}+ \frac{Z}{r_2}-\frac{p^2_2}{2} \bigg)4\pi Z\delta^3(\vec{r}_1) -\frac{5 Z}{4}\frac{\vec{r}}{r^3}\cdot\bigg(\frac{\vec{r}_1}{r^3_1}-\frac{\vec{r}_2}{r^3_2} \bigg)+ 2\frac{Z}{r_1} p_2^i\bigg(\frac{\delta^{ij}}{r}+\frac{r^ir^j}{r^3}\bigg)p^j_1 \bigg\rangle \\
&- \frac{2}{3}\bigg\langle \frac{Z}{r_1} +\frac{Z}{r_2} \bigg\rangle \bigg\langle \bigg(E-\frac{1}{r}+ \frac{Z}{r_1}+ \frac{Z}{r_2} \bigg)^2- \frac{p^2_1p^2_2}{2}+ p_2^i\bigg(\frac{\delta^{ij}}{r}+\frac{r^ir^j}{r^3}\bigg)p^j_1 \bigg \rangle. \\
\end{aligned}
\end{equation}
\end{widetext}

The second-order $m\alpha^6$ corrections $E^{(6)}_{\mathrm{sec}} \equiv \langle H_{\mathrm{hfs}}^{(4)} , [H_{\mathrm{nfs}}^{(4)}+H_{\mathrm{fs}}^{(4)}] \rangle$ and $E^{(6)}_{\mathrm{hfs}} \equiv  \langle H_{\mathrm{hfs}}^{(4)} , H_{\mathrm{hfs}}^{(4)} \rangle _{\mathrm{hfs}}$ are easy to understand, but due to the coupling with the intermediate states of different symmetries and the presence of singular operators, their calculations are enormously challenging. For $E^{(6)}_{\rm{sec}}$, we eliminate the singularities by adding a singular term to the wave function in each intermediate state, and then take all possible intermediate states into consideration. The calculation of the second-order hyperfine correction $E^{(6)}_{\rm{hfs}}$ is difficult. Fortunately, since $2\,^1\!P_1-2\,^3\!P_J$ is quite small, to a good approximation, we can only include the dominant contribution from the $2\,^1\!P_1$ intermediate state.

\section{QED and nuclear corrections}\label{sec2}
This section will discuss the corrections due to QED and nuclear effects. The effective operator for the QED correction of order $m\alpha^6$ is
\begin{equation}\label{eq:27}
    \begin{aligned}
        H_{\mathrm{QED}}^{(6)} =Z\alpha^2 \bigg( \ln2-\frac{5}{2} \bigg)  H^Q_{\rm{hfs}},  \\
    \end{aligned}
\end{equation}
where $H^Q_{\rm{hfs}} = C^1_{34}  \frac{4\pi Z}{3}  [\delta^3(\vec{r}_1) + \delta^3(\vec{r}_2)](\vec{I} \cdot \vec{S})$. For the QED corrections of order $m\alpha^7$, we use the hydrogenic approximation~\cite{Karshenboim2002,Puchalski2013}, {\it i.e.},
\begin{equation}\label{eq:28}
    \begin{aligned}
        H_{\mathrm{QED}}^{\mathrm{ho}} (1s) =& \bigg\{\frac{\alpha}{\pi} (Z \alpha)^2 \bigg[ -\frac{8}{3} \ln Z \alpha (\ln Z \alpha - 0.8009) + 16.9038 \bigg] + 0.771652 \frac{\alpha^2}{\pi} (Z \alpha) \bigg\}  H^Q_{\rm{hfs}}\,,\\
    \end{aligned}
\end{equation}
and
\begin{equation}\label{eq:29}
    \begin{aligned}
        H_{\mathrm{QED}}^{\mathrm{ho}} (2s) =& \bigg\{\frac{\alpha}{\pi} (Z \alpha)^2 \bigg[ -\frac{8}{3} \ln Z \alpha (\ln Z \alpha + 0.4378 ) + 11.3522 \bigg] + 0.771652 \frac{\alpha^2}{\pi} (Z \alpha) \bigg\}  H^Q_{\rm{hfs}}\,.\\
    \end{aligned}
\end{equation}
We may further adopt the following weighted approximation~\cite{McKenzieR1991,Yan2000}
\begin{equation}\label{eq:30}
    \begin{aligned}
        H_{\mathrm{QED}}^{\mathrm{ho}} (1s2s) = \frac{H_{\mathrm{QED}}^{\mathrm{ho}} (1s) + H_{\mathrm{QED}}^{\mathrm{ho}} (2s)/8 }{1+1/8}\,.
    \end{aligned}
\end{equation}
The nuclear contribution due to the Zemach term is
\begin{equation}\label{eq:31}
\begin{aligned}
H_{\mathrm{nucl}} = -2ZR_{\mathrm{em}}  H^Q_{\rm{hfs}},
\end{aligned}
\end{equation}
where the Zemach radius $R_{\mathrm{em}}$ is the average electromagnetic charge radius of the nucleus
\begin{equation}\label{eq:32}
\begin{aligned}
R_{\mathrm{em}} = \int d^3r d^3r' \rho_e(r) \rho_m(r') |\vec{r}-\vec{r}'|.
\end{aligned}
\end{equation}
The contribution of the nuclear electric quadrupole moment $Q_d$ is related to the nuclear spin-spin tensor operator~\cite{Zhang2013}
\begin{equation}\label{eq:33}
H_{\mathrm{eqm}} = \frac{Q_d}{2} I^iI^j \bigg[\frac{1}{r_1^3} \bigg( \delta^{ij} - 3\frac{r_1^ir_1^j}{r_1^2} \bigg)
+\frac{1}{r_2^3} \bigg( \delta^{ij} - 3\frac{r_2^ir_2^j}{r_2^2} \bigg) \bigg],
\end{equation}
where $Q_d$ is $-0.000806(6)\times 10^{-24}$~cm$^2$ for $^6\mathrm{Li}^+$ and $-0.0400(3)\times 10^{-24}$~cm$^2$ for $^7\mathrm{Li}^+$~\cite{Stone2016}.

\section{Numerical results for the $2\,^3\!S_1$ state}\label{sec3}

For the expectation values of $m\alpha^4$ and $m\alpha^6$ operators, only the scalar operators are nonzero for $2\,^3\!S_1$, as listed in
Table~\ref{tab:1}.
The second-order correction $E^{(6)}_{\mathrm{sec}}$ of $m\alpha^6$ can be divided into three parts according to the symmetries of the intermediate states $^3\!S$, $^3\!P$, and $^3\!D$. For the $2\,^3\!S_1$ state, the only nonzero angular momentum coefficients are $I^iS^i$ and $\{I^iI^j\}\{S^iS^j\}$.
\begin{table}[ht]
    \caption{\label{tab:1} Expectation values for the $2\,^3\!S_1$ states of $^6$Li$^+$ and $^7$Li$^+$. The listed numerical values are uncertain only at the last digits. In atomic units.}
     \begin{ruledtabular}
     \begin{tabular}{lcc}
     \multicolumn{1}{c}{Operator}
    &\multicolumn{1}{c}{$^6\mathrm{Li}^+$}
    &\multicolumn{1}{c}{$^7\mathrm{Li}^+$}  \\
    \hline
    $4\pi \delta^3(\vec{r}_1)$   &\,\,57.350361  &\,\,57.350360  \\
    $K'$                   &--22.396754     &--22.396754     \\
    \end{tabular}
    \end{ruledtabular}
\end{table}
Thus, we have
\begin{equation}\label{eq:34}
\begin{aligned}
E^{(6)}_{\mathrm{sec}} (2\,^3\!{S_1}) = C^1_{36} \big[E^{(6)}_{^3\!{S}} (2\,^3\!{S_1})+E^{(6)}_{^3\!{P}} (2\,^3\!{S_1})+ E^{(6)}_{^3\!{D}} (2\,^3\!{S_1})\big],
\end{aligned}
\end{equation}
where
\begin{equation}\label{eq:35}
\begin{aligned}
E^{(6)}_{^3\!{S}} (2\,^3\!{S_1}) &=\langle P',G' \rangle \langle 2I^iS^i \rangle,
\end{aligned}
\end{equation}
\begin{equation}\label{eq:36}
\begin{aligned}
E^{(6)}_{^3\!{P}} (2\,^3\!{S_1}) &= \langle \vec{P},\vec{G} \rangle \bigg\langle \frac{2}{3}I^iS^i \bigg\rangle,
\end{aligned}
\end{equation}
\begin{equation}\label{eq:37}
\begin{aligned}
E^{(6)}_{^3\!{D}} (2\,^3\!{S_1}) &= \langle \hat{P},\hat{G} \rangle \bigg\langle \frac{1}{3}I^iS^i \bigg\rangle.
\end{aligned}
\end{equation}
For $E^{(6)}_{\mathrm{hfs}}$, we have
\begin{equation}\label{eq:38}
\begin{aligned}
E^{(6)}_{\mathrm{hfs},^1\!{S}} (2\,^3\!{S_1}) &\approx C^2_{56} \langle P_A,P_A \rangle^\circ \bigg\langle -\frac{1}{2}I^iS^i-\{I^iI^j\}\{S^iS^j\} \bigg\rangle,
\end{aligned}
\end{equation}
where $\langle A,B \rangle^\circ \equiv \frac{\langle 2\,^3\!\vec{\chi} |A| 2\,^1\!\vec{\chi} \rangle\langle 2\,^1\!\vec{\chi} |B| 2\,^3\!\vec{\chi} \rangle}{E(2\,^3\!\chi)-E(2\,^1\!\chi)} $. Numerical results for the radial parts in $E^{(6)}_{\mathrm{sec}}$ and $E^{(6)}_{\mathrm{hfs}}$ are presented in Table~\ref{tab:2}.
\begin{table}[ht]
    \caption{\label{tab:2} Individual second-order matrix elements for all possible intermediate states connected to $2\,^3\!S_1$.
      The listed numerical values are uncertain only at the last digits. In atomic units.
     }
     \begin{ruledtabular}
     \begin{tabular}{lcc}
     \multicolumn{1}{c}{Intermediate state}
    &\multicolumn{1}{c}{$\langle A, B \rangle$}
    &\multicolumn{1}{c}{Value} \\
    \hline
    $^3S$   &$\langle P',G' \rangle$                 & 799.59                              \\
    $^3P$   &$\langle \vec{P},\vec{G} \rangle$       & 0.0535                        \\
    $^3D$   &$\langle \hat{P},\hat{G} \rangle$       & 0.040(5)                      \\
    $^1S$   &$\langle P_A,P_A \rangle^\circ$         &--136622.260                                      \\
    \end{tabular}
    \end{ruledtabular}
\end{table}

\section{Numerical results for the $2\,^3\!P_J$ state}\label{sec4}

\begin{table}[ht]
    \caption{\label{tab:4} Expectation values for the $2\,^3\!P_J$ states of $^6$Li$^+$ and $^7$Li$^+$. The listed numerical values are uncertain only at the last digits. In atomic units.}
     \begin{ruledtabular}
     \begin{tabular}{lcc}
     \multicolumn{1}{c}{$ $}
    &\multicolumn{1}{c}{$^6\mathrm{Li}^+$}
    &\multicolumn{1}{c}{$^7\mathrm{Li}^+$} \\
    \hline
    $4\pi \delta^3(\vec{r}_1)$                                         &53.219006    &53.219058    \\
    $(\vec{r}_1 \times \vec{p}_1)/r_1^3$                               &0.479843     &0.479824   \\
    $(\vec{r}_1 \times \vec{p}_2)/r_1^3$                               &--0.691221   &--0.691199  \\
    $(\delta^{ij} - 3r_1^ir_1^j/r_1^2)/r_1^3$                          &--0.487622   &--0.487606  \\
    $K'$                                                               &7.805484     &7.805484   \\
    $\vec{K}$                                                          &--15.742681  &--15.742681       \\
    $\hat{K}$                                                          &--10.371686  &--10.371686       \\
    \end{tabular}
    \end{ruledtabular}
\end{table}

The expectation values for the operators of $m\alpha^4$ and $m\alpha^6$ can be evaluated in a similar way as $2\,^3\!S_1$. The results are listed in Table~\ref{tab:4}.
The second-order correction $E^{(6)}_{\mathrm{sec}}$ of $m\alpha^6$ can be similarly divided into five parts according to the symmetries of the
intermediate states $^3\!P$, $^1\!P$, $^3\!D$, $^1\!D$, and $^3\!F$:
\begin{equation}\label{eq:39}
E^{(6)}_{\mathrm{sec}} (2\,^3{P_J}) = C^1_{36} \big[E^{(6)}_{^3{P}} (2\,^3{P_J})+(1+a_e)E^{(6)}_{^1{P}} (2\,^3{P_J})
+ E^{(6)}_{^3{D}} (2\,^3{P_J}) + E^{(6)}_{^1{D}} (2\,^3{P_J}) + E^{(6)}_{^3{F}} (2\,^3{P_J}) \big].
\end{equation}
For $2\,^3\!P_J$, the angular momentum operators that are relevant to $E^{(6)}_{\mathrm{sec}}$ are
$I^iS^i$, $I^iL^i$, $I^iS^j\{L^iL^j\}$, and $I^iL^j\{S^iS^j\}$;
for $E^{(6)}_{\mathrm{hfs}}$, however, they are $I^iS^i$, $I^iL^i$, $I^iS^j\{L^iL^j\}$, $I^iL^j\{S^iS^j\}$, $\{I^iI^j\}\{S^iS^j\}$, $\{I^iI^j\}\{L^iL^j\}$, $S^iL^j\{I^iI^j\}$, and $\{I^iI^j\}\{\{S^mS^n\}\{L^kL^l\}\}^{ij}$.
The individual contributions in Eq.~(\ref{eq:39}) are given by
\begin{widetext}
\begin{equation}\label{eq:40}
\begin{aligned}
E^{(6)}_{^3{P}} (2\,^3{P_J}) &=\langle P',G' \rangle \langle 2I^iS^i \rangle + \langle \vec{P},G \rangle \langle I^iL^i \rangle + \langle \hat{P},G \rangle \bigg\langle -\frac{6}{5}I^iS^j\{L^iL^j\} \bigg\rangle \\
&+ \langle P,\vec{G} \rangle \bigg\langle \frac{2}{3}I^iL^i+I^iL^j\{S^iS^j\}\bigg\rangle + \langle P,\hat{G} \rangle \bigg\langle -\frac{3}{5}I^iS^j\{L^iL^j\} \bigg\rangle \\
&+ \langle \vec{P},\vec{G} \rangle \bigg\langle \frac{1}{3}I^iS^i+\frac{1}{2}I^iS^j\{L^iL^j\} \bigg\rangle + \langle \vec{P},\hat{G} \rangle \bigg\langle -\frac{3}{10}I^iL^j\{S^iS^j\} \bigg\rangle \\
&+ \langle \hat{P},\vec{G} \rangle \bigg\langle -\frac{1}{3}I^iL^i+\frac{9}{20}I^iS^j\{L^iL^j\}-\frac{1}{20}I^iL^j\{S^iS^j\} \bigg\rangle \\
&+ \langle \hat{P},\hat{G} \rangle \bigg\langle \frac{1}{5}I^iS^i-\frac{21}{100}I^iS^j\{L^iL^j\}-\frac{27}{200}I^iL^j\{S^iS^j\} \bigg\rangle, \\
\end{aligned}
\end{equation}
\begin{equation}\label{eq:41}
\begin{aligned}
E^{(6)}_{^1{P}} (2\,^3{P_J}) &=\langle P_A,\vec{G}_A \rangle \bigg\langle \frac{1}{3}I^iL^i-I^iL^j\{S^iS^j\} \bigg\rangle \\
&+ \langle \hat{P}_A,\vec{G}_A \rangle \bigg\langle -\frac{1}{6}I^iL^i+\frac{9}{20}I^iS^j\{L^iL^j\}+\frac{1}{20}I^iL^j\{S^iS^j\} \bigg\rangle, \\
\end{aligned}
\end{equation}
\begin{equation}\label{eq:42}
\begin{aligned}
E^{(6)}_{^3{D}} (2\,^3{P_J}) &= \langle \vec{P},\vec{G} \rangle \bigg\langle \frac{2}{3}I^iS^i-\frac{1}{5}I^iS^j\{L^iL^j\} \bigg\rangle + \langle \vec{P},\hat{G} \rangle \bigg\langle -\frac{3}{5}I^iL^j\{S^iS^j\} \bigg\rangle \\
&+ \langle \hat{P},\vec{G} \rangle \bigg\langle -\frac{2}{3}I^iL^i-\frac{3}{10}I^iS^j\{L^iL^j\}-\frac{1}{10}I^iL^j\{S^iS^j\} \bigg\rangle \\
&+ \langle \hat{P},\hat{G} \rangle \bigg\langle \frac{2}{9}I^iS^i+\frac{7}{30}I^iS^j\{L^iL^j\}-\frac{1}{10}I^iL^j\{S^iS^j\} \bigg\rangle,\\
\end{aligned}
\end{equation}
\begin{equation}\label{eq:43}
\begin{aligned}
E^{(6)}_{^1{D}} (2\,^3{P_J}) &= \langle \hat{P}_A,\vec{G}_A \rangle \bigg\langle -\frac{1}{3}I^iL^i-\frac{3}{10}I^iS^j\{L^iL^j\}+\frac{1}{10}I^iL^j\{S^iS^j\} \bigg\rangle, \\
\end{aligned}
\end{equation}
\begin{equation}\label{eq:44}
\begin{aligned}
E^{(6)}_{^3{F}} (2\,^3{P_J}) &= \langle \hat{P},\hat{G} \rangle \bigg\langle \frac{1}{3}I^iS^i-\frac{1}{10}I^iS^j\{L^iL^j\}+\frac{3}{10}I^iL^j\{S^iS^j\} \bigg\rangle. \\
\end{aligned}
\end{equation}
The second-order hyperfine correction is
\begin{equation}\label{eq:45}
\begin{aligned}
E^{(6)}_{\mathrm{hfs},^1{P}} (2\,^3{P_J}) &\approx C^2_{56} \bigg\{ \langle P_A,P_A \rangle^\circ \bigg\langle -\frac{1}{2}I^iS^i-\{I^iI^j\}\{S^iS^j\} \bigg\rangle \\&+ \langle P_A,\hat{P}_A \rangle^\circ \bigg\langle -\frac{3}{10}I^iS^j\{L^iL^j\}-\frac{2}{5}\{I^iI^j\}\{L^iL^j\}-\frac{6}{5}\{I^iI^j\}\{\{S^mS^n\}\{L^kL^l\}\}^{ij} \bigg\rangle \\
&+ \langle \hat{P}_A,\hat{P}_A \rangle^\circ \bigg\langle \frac{1}{20}I^iS^i-\frac{3}{40}I^iL^i+\frac{21}{200}I^iS^j\{L^iL^j\}-\frac{9}{200}I^iL^j\{S^iS^j\}-\frac{1}{50}\{I^iI^j\}\{S^iS^j\} \\
&\quad\quad\quad\quad\quad\quad -\frac{7}{100}\{I^iI^j\}\{L^iL^j\}+\frac{9}{200}S^iL^j\{I^iI^j\}+\frac{3}{50}\{I^iI^j\}\{\{S^mS^n\}\{L^kL^l\}\}^{ij} \bigg\rangle \bigg\}.
\end{aligned}
\end{equation}
\end{widetext}
Numerical results for the radial parts of $E^{(6)}_{\mathrm{sec}}$ and $E^{(6)}_{\mathrm{hfs}}$ are presented in Table~\ref{tab:5}.

\begin{table}[ht]
    \caption{\label{tab:5}
    Individual second-order matrix elements for all possible intermediate states connected to $2\,^3\!P_J$.
    The listed numerical values are uncertain only at the last digits. In atomic units.
    }
     \begin{ruledtabular}
     \begin{tabular}{lcc}
     \multicolumn{1}{c}{Intermediate state}
    &\multicolumn{1}{c}{$\langle A, B \rangle$}
    &\multicolumn{1}{c}{Value} \\
    \hline
    $^3P$   &$\langle P',G' \rangle$                       & 717.56                 \\
            &$\langle \vec{P},G \rangle$                   & 8.69             \\
            &$\langle \hat{P},G \rangle$                   & 4.08(5)          \\
            &$\langle P,\vec{G} \rangle$                   & 3.63             \\
            &$\langle \vec{P},\vec{G} \rangle$             & --5.79                \\
            &$\langle \hat{P},\vec{G} \rangle$             & --4.21                \\
            &$\langle P,\hat{G} \rangle$                   & 0.674            \\
            &$\langle \vec{P},\hat{G} \rangle$             & 3.584            \\
            &$\langle \hat{P},\hat{G} \rangle$             & 1.674            \\
    $^1P$   &$\langle P_A,\vec{G}_A \rangle$               & 2201.827                  \\
            &$\langle \hat{P}_A,\vec{G}_A \rangle$         & --33.30(5)               \\
    $^3D$   &$\langle \vec{P},\vec{G} \rangle$             & 0.011            \\
            &$\langle \hat{P},\vec{G} \rangle$             & --0.004               \\
            &$\langle \vec{P},\hat{G} \rangle$             & 0.281            \\
            &$\langle \hat{P},\hat{G} \rangle$             &--0.011               \\
    $^1D$   &$\langle \hat{P}_A,\vec{G}_A \rangle$         &0.0546           \\
    $^3F$   &$\langle \hat{P},\hat{G} \rangle$             & 0.111            \\
    $^1P$   &$\langle P_A,P_A \rangle^\circ$               &--332509.770                         \\
            &$\langle P_A,\hat{P}_A \rangle^\circ$         &13923.581                    \\
            &$\langle \hat{P}_A,\hat{P}_A \rangle^\circ$   &--158.510                   \\
    \end{tabular}
    \end{ruledtabular}
\end{table}

\section{Singlet-triplet mixing}\label{sec5}
We compare two different methods used in dealing with the $2\,^1\!P_1-2\,^3\!P_1$ mixing effect. Method 1. Do an exact diagonalization only within the $2\,^3\!P_J$ manifold and treat the $2\,^1\!P_1-2\,^3\!P_1$ mixing effect by perturbation theory up to second order. Method 2. Extend the
$2\,^3\!P_J$ manifold by including the $2\,^1\!P_1$ state and do an exact diagonalization of the extended matrix. Both the methods only include the relativistic correction
of order $m\alpha^4$.

Using standard second-order perturbation theory within the $2\,^3\!P_1$, $2\,^1\!P_1$ basis set, and for purposes of illustration without hyperfine structure, the $E(2\,^3\!P_1)$ energy can be written in the form
\begin{equation}\label{eq:46}
\begin{aligned}
E(2\,^3\!P_1)=E_0+X,
\end{aligned}
\end{equation}
where $E_0$ is the energy without the $2\,^1\!P_1-2\,^3\!P_1$ mixing, and $X$ is the contribution from the $2\,^1\!P_1-2\,^3\!P_1$ mixing calculated by perturbation theory
\begin{equation}\label{eq:47}
    \begin{aligned}
     X = \frac{|\langle 2\,^3\!P |H_{\mathrm{mix}}| 2\,^1\!P \rangle|^2}{E(2\,^3\!P)-E(2\,^1\!P)}.
    \end{aligned}
\end{equation}
However, with increasing $Z$ or $L$, higher-order perturbation corrections become important because the energy denominator is relatively small for this term. This can be taken into account by subtracting $X$ out and replacing it with the energy shift $\Delta$ obtained by an exact diagonalization of the $2\times 2$ Hamiltonian in the $2\,^1\!P_1$, $2\,^3\!P_1$ basis set.  The result for Method 2 is
\begin{equation}\label{eq:48}
    \begin{aligned}
    \tilde{E}(2\,^3P_1) = E_0 + X - \frac{|\langle 2\,^3\!P |H_{\mathrm{mix}}| 2\,^1\!P \rangle|^2}{E(2\,^3\!P)-E(2\,^1\!P)} + \Delta,
    \end{aligned}
\end{equation}
Therefore, with the identification $E = E_0 + X$, Eq.(4) in the paper can be written in the form
\begin{equation}\label{eq:49}
    \begin{aligned}
    \tilde{E} = E - X + \Delta.
    \end{aligned}
\end{equation}
Thus, with hyperfine structure included in an extended diagonalization, the energy shift between levels 1 and 2 is
\begin{equation}\label{eq:50}
    \begin{aligned}
    \Delta \tilde{E}_{12} = \Delta E_{12} - (X_1-X_2) + \Delta_1-\Delta_2.
    \end{aligned}
\end{equation}
We defined the difference between these two methods as
\begin{equation}\label{eq:51}
    \begin{aligned}
    \delta = - (X_1-X_2) + \Delta_1-\Delta_2.
    \end{aligned}
\end{equation}

The second-order matrix elements involving the intermediate state $2\,^1\!P_1$ and the hyperfine structure coefficients for the $2\,^1\!P_1$ and $2\,^3\!P_J$ states are listed in Tables~\ref{tab:10} and \ref{tab:9}, as inputs for applying Methods 1 and 2, respectively.
The uncertainty of the Method 2 is due mainly to the uncertainty in the off-diagonal singlet-triplet matrix elements~\cite{Riis1994}. The hyperfine splittings in $2\,^3\!P_J$ of $^{6,7}$Li$^+$ were evaluated using these two methods and the results are presented in Tables~\ref{tab:11} and \ref{tab:12}, respectively.

\begin{table}[ht]
 \caption{\label{tab:10} Numerical values of second-order matrix elements for the intermediate state $2\,^1\!P_1$.
    The listed numerical values are uncertain only at the last digits. In atomic units.
    }
     \begin{ruledtabular}
     \begin{tabular}{lc}
    \multicolumn{1}{l}{$\langle A, B \rangle$}
    &\multicolumn{1}{c}{Value} \\
    \hline
    $\langle P_A,\vec{G}_A \rangle^\circ$          &2184.476    \\
    $\langle \hat{P}_A,\vec{G}_A \rangle^\circ$    &--24.839     \\
    \end{tabular}
    \end{ruledtabular}
\end{table}

\begin{table}[ht]
\caption{\label{tab:9} Calculated values of the $^6$Li$^+$ and $^7$Li$^+$ hyperfine structure coefficients for the $2\,^1\!P_1$ and
    $2\,^3\!P_J$ states, in MHz. These coefficients are defined in Eqs.~(10)-(12) of Ref.~\cite{Riis1994}.
    The listed numerical values are uncertain only at the last digits.
    }
     \begin{ruledtabular}
     \begin{tabular}{lcc}
     \multicolumn{1}{c}{Coefficient.}
    &\multicolumn{1}{c}{$^6$Li$^+$}
    &\multicolumn{1}{c}{$^7$Li$^+$}\\
    \hline
     $C^{(0)}_{1,1}$ &2782.717       &7348.908            \\
     $C^{(0)}_{1,0}$ &2794.668       &7380.461            \\
     $D^{(0)}_{1}$   &37.635         &99.387         \\
     $D^{(0)}_{0}$   &24.379         &64.383         \\
     $E^{(0)}_{1,1}$ &--9.561        &--25.250     \\
     $E^{(0)}_{1,0}$ &--7.945        &--20.981     \\
    \end{tabular}
    \end{ruledtabular}
\end{table}

\begin{table}[ht]
\caption{\label{tab:11} Hyperfine splittings in $2\,^3\!P_J$ of $^6$Li$^+$, in MHz. The listed numerical values are uncertain only at the last digits.
    }
     \begin{ruledtabular}
     \begin{tabular}{cccc}
     $(J,F)-(J',F')$ &Method 1  &Method 2  &$\delta$  \\
    \hline
     $(2,1)-(2,2)$ &2855.068  &2855.068           &0.000           \\
     $(2,2)-(2,3)$ &4123.962  &4123.962           &0.000           \\
     $(1,0)-(1,1)$ &1316.542  &1316.543(2)        &0.001(2) \\
     $(1,1)-(1,2)$ &2885.438  &2885.440(2)        &0.002(2) \\
    \end{tabular}
    \end{ruledtabular}
\end{table}
\begin{table}[ht]
    \caption{\label{tab:12} Hyperfine splittings in $2\,^3\!P_J$ of $^7$Li$^+$, in MHz. The listed numerical values are uncertain only at the last digits.
       }
     \begin{ruledtabular}
     \begin{tabular}{cccc}
     $(J,F)-(J',F')$      &Method 1        &Method 2            &$\delta$             \\
    \hline
     $(2,1/2)-(2,3/2)$  &6195.554  &6195.553         &--0.001               \\
     $(2,3/2)-(2,5/2)$  &9597.308  &9597.309         &0.001            \\
     $(2,5/2)-(2,7/2)$  &11765.303       &11765.308              &0.005            \\
     $(1,1/2)-(1,3/2)$  &4238.158  &4238.172(2) &0.014(2)  \\
     $(1,3/2)-(1,5/2)$  &9956.074  &9956.085(3) &0.011(3)  \\
    \end{tabular}
    \end{ruledtabular}
\end{table}

\clearpage

%
